\documentstyle[12pt]{article}
\begin{document}
\baselineskip=15.5pt
\begin{titlepage}

\begin{flushright}
IC/2001/4\\
hep-th/0101228
\end{flushright}
\vspace{10 mm}

\begin{center}
{\Large Brane World Cosmologies with Varying Speed of Light}

\vspace{5mm}

\end{center}

\vspace{5 mm}

\begin{center}
{\large Donam Youm\footnote{E-mail: youmd@ictp.trieste.it}}

\vspace{3mm}

ICTP, Strada Costiera 11, 34014 Trieste, Italy

\end{center}

\vspace{1cm}

\begin{center}
{\large Abstract}
\end{center}

\noindent

We study cosmologies in the Randall-Sundrum models, incorporating 
the possibility of time-varying speed of light and Newton's constant.  
The cosmologies with varying speed of light (VSL) were proposed by 
Moffat and by Albrecht and Magueijo as an alternative to inflation for 
solving the cosmological problems.  We consider the case in which the 
speed of light varies with time after the radion or the scale of the 
extra dimension has been stabilized.  We elaborate on the conditions 
under which the flatness problem and the cosmological constant problem 
can be resolved.  We find that the RS models are more restrictive about 
possible desirable VSL cosmological models than the standard general 
relativity.  Particularly, the VSL cosmologies may provide with a possible 
mechanism for bringing the quantum corrections to the fine-tuned brane 
tensions after the SUSY breaking under control.

\vspace{1cm}
\begin{flushleft}
January, 2001
\end{flushleft}
\end{titlepage}
\newpage

\section{Introduction}

Although successful in passing some crucial observational tests, the 
Standard Big Bang (SBB) model is regarded as being incomplete in the sense 
that it does not address the state of universe at initial period and many 
features of the SBB model are just assumed as initial conditions.  As a 
consequence, the SBB model has many cosmological problems which the SBB model 
can avoid only by assuming unnatural or fine-tuned initial conditions.  The 
inflationary scenario \cite{gut,lin,als} was introduced in an attempt to 
complete the SBB picture as its precursor and solve some of the cosmological 
problems.  The inflationary scenario assumes a period of superluminary 
or accelerated expansion during an initial period (prior to a conventional 
non-inflationary evolution described by the SBB model) due to a scalar field 
(inflaton) with an inflationary potential satisfying the slow-roll 
approximation.  The inflationary models successfully solve the horizon 
problem, the flatness problem, the homogeneity problem, and the problem with 
a variety of unwanted relics without assuming unnatural initial conditions.  
The inflationary models also have been favored as prominent candidates for 
explaining the origin of structure in the universe.

The varying speed of light (VSL) theory was proposed \cite{mof1,am} as an 
alternative to inflation for solving the cosmological problems.  Instead of 
introducing an inflaton to make the universe undergo superluminal expansion 
while violating the strong energy condition, the VSL models just assume 
without changing the universe matter content that the speed of light initially 
took a larger value and then decreased to the present value at an early time, 
thereby violating the Lorentz invariance.  As the speed of light is assumed 
to have taken a larger value at an early time, the horizon problem is 
automatically solved.  The VSL models also solve 
\cite{mof1,am,bar1,bar2,bar3,mof2,bar4} other various cosmological problems 
that the inflationary models solve.  What is appealing about the VSL models 
is that unlike the inflationary models the cosmological constant problem 
can be solved if the rapid enough decreasing speed of light is assumed.  
The claim in Ref. \cite{wfc} of the experimental evidence for a 
time-varying fine structure constant $\alpha=e^2/(4\pi\hbar c)$ suggests 
that the speed of light may indeed vary with time.  (Other possibility 
of time-varying electric charge $e$ was considered in Ref. \cite{bek}.)  
However, in terms of theoretical foundation, the VSL models are not yet 
as well developed as the inflationary ones.  (Some of attempts can be 
found in Refs. \cite{mof1,mof3,man,dru,ste,mof4,mof5,mag}.)  

In the brane world scenario with warped compactification \cite{rs1,rs2,rs3}, 
it is shown that the light signal can travel faster through the extra 
dimension and thereby the horizon problem can be solved 
\cite{kal1,kal2,chu,ish}.  Further aspect of the Lorentz violation, which 
is a necessary requirement (Cf. Ref. \cite{ave}) for the VSL models, 
was explored in Ref. \cite{ckr,csa}, where the possibility of gravitons, 
instead of photons, traveling faster is considered.  Ref. \cite{kir} gave 
a toy example on the brane world cosmology where the speed of light varies 
with time.  (See also Ref. \cite{ale} for further study on this aspect.)  
So, the brane world scenario may provide with a natural framework or 
mechanism for realizing the VSL cosmology.  Moreover, since the VSL models 
solve the cosmological constant problem, brane world cosmologies with 
variable speed of light may be used to control the quantum corrections to 
the fine-tuned brane tensions after the SUSY breaking, just like the 
mechanism for self-tuning brane tension \cite{adk,kss}:  While the speed of 
light changes with time, the quantum corrections to the brane tensions 
get converted into ordinary matter, thereby the brane tensions being pushed 
back to the fine-tuned values.  

In this paper, we study the VSL cosmologies in the Randall-Sundrum (RS) 
models.  We consider the case in which the speed of light changes with 
time after the radion or the scale of the extra dimension has been 
stabilized.  Following the previous works \cite{mof1,am,bar1}, we just 
assume the speed of light to change either suddenly at some critical time 
or gradually as a power law in the cosmic scale factor without specifying 
the mechanism for generating time-variable speed of light, for the purpose 
of seeing consequences for varying speed of light in the brane world 
cosmologies.  It is a subject of future research to find natural mechanism 
for realizing the time-varying speed of light within realistic brane world 
cosmologies.  We elaborate on conditions under which the flatness problem 
and the cosmological constant problem can be resolved.  

The paper is organized as follows.  In section 2, we study the VSL 
cosmology in the RS model with one positive tension and one negative 
tension branes (RS1 model).  We consider the cases of sudden and 
gradual changes of the speed of light with time.  In section 3, we 
study the VSL cosmology in the RS model with one positive tension 
brane (RS2 model) for the case when the speed of light suddenly 
changes with time.

\section{The VSL Cosmology in the RS1 Model}

In this section, we study the cosmology in the RS model with one positive 
tension and one negative tension branes, considering the possibility of 
time-varying speed of light and five-dimensional Newton's constant.  We 
place the brane with positive tension $\sigma_0$ [negative tension 
$\sigma_{1/2}$] at $y=0$ [at $y=1/2$].  The bulk contains a cosmological 
constant $\Lambda$.  We add a massive bulk scalar $\Phi$ with potentials 
$V_{0,1/2}(\Phi)$ on the branes at $y=0,1/2$, having nontrivial VEV, to 
stabilize the radion through the Goldberger-Wise mechanism \cite{gw,gw1,gw2}. 
The expansion of the brane universe is due to matter fields with the 
Lagrangian densities ${\cal L}_{0,1/2}$ on the branes at $y=0,1/2$.  

We assume that the speed of light $c$ varies with time during some initial 
period of cosmological evolution.  $c$ may go through a first order phase 
transition at a critical time $t_c$, suddenly jumping from some superluminal 
value $c_-$ to the present value $c_+$ of the speed of light, as was 
considered in Refs. \cite{mof1,am}, or may change gradually like 
$c(t)=c_0a^{\sf n}$ for some constants $c_0$ and ${\sf n}$ \cite{bar1}.  
In Ref. \cite{mof1}, it is proposed that the superluminary phase of the 
former case may be due to the spontaneous Higgs breaking of local Lorentz 
invariance from $SO(3,1)$ to $SO(3)$ above some critical temperature $T_c$.  
In general, if $c$ is variable, then the Lorentz invariance becomes 
explicitly broken.  It is postulated in VSL models that there exists a 
preferred Lorentz frame in which the laws of physics simplify.  In such 
preferred frame, the action is assumed to be given by the standard (Lorentz 
invariant) action with a constant $c$ replaced by a field $c(x^{\mu})$, 
{\it the principle of minimal coupling}.  However, such simplified laws of 
physics are not invariant under the frame transformation.  

So, the action under consideration is given by
\begin{eqnarray}
S&=&\int d^5x\left[\sqrt{-\hat{g}}\left({\psi\over{16\pi G_5}}{\cal R}-\Lambda
-{1\over 2}\partial_M\Phi\partial^M\Phi+{1\over 2}m^2\Phi^2\right)
+{\cal L}_{\psi}\right]
\cr
&+&\int d^4x\sqrt{-g}\left[{\cal L}_0+\left.V_0(\Phi)\right|_0-\sigma_0\right]
+\int d^4x\sqrt{-g}\left[{\cal L}_{1/2}+\left.V_{1/2}(\Phi)\right|_{1/2}
-\sigma_{1/2}\right],
\label{sgbrnact}
\end{eqnarray}
where in analogy with the Brans-Dicke theory we defined a scalar field 
$\psi\equiv c^4$ out of varying speed of light $c$, $G_5$ is the 
five-dimensional Newton's constant
\footnote{With our convention of the D-dimensional Newton's constant $G_D$, 
i.e. $\kappa_D=8\pi G_D/c^4$, which most of papers follow, the Newton's 
law of gravitation, which is reproduced in the weak gravitational field 
and low velocity limit of the Einstein's gravity, takes the form 
$F_{12}={{4\pi G_D}\over{\Omega_{D-2}}}{{m_1m_2}\over{r^{D-2}}}$, where 
$\Omega_{D-2}$ is the volume of the unit $(D-2)$-sphere, instead of the 
standard form $F_{12}=G_D{{m_1m_2}\over{r^{D-2}}}$.}, 
and ${\cal L}_{\psi}$ controls the dynamics of $\psi$.  
It is required that ${\cal L}_{\psi}$ should be explicitly independent of 
the other fields, including the metric, so that the principle of minimal 
coupling continues to hold for the field equations.  This requirement 
implies that Bianchi identities ${\cal G}^{\mu\nu}_{\ \ \ ;\nu}=0$ of the 
Einstein tensor ${\cal G}_{\mu\nu}$ continue to hold and thereby the 
energy-momentum conservation law is modified by an additional term 
proportional to $\dot{c}/c$.  (Cf. See the Einstein's equations (\ref{eineq}) 
in the below.)  This violation of the energy-momentum tensor conservation 
during the phase of varying speed of light could provide with an 
explanation for the creation of matter in the beginning of universe.  This 
matter creation process is somehow analogous to the reheating process of the 
inflationary models, by which the inflaton's energy density is converted to 
conventional matter after inflation.  

As usual, the metric Ansatz for the expanding brane universe is given by
\begin{equation}
\hat{g}_{MN}dx^Mdx^N=-n^2(t,y)c^2dt^2+a^2(t,y)\gamma_{ij}dx^idx^j+b^2(t,y)dy^2,
\label{metansat}
\end{equation}
where $\gamma_{ij}$ is the metric for the maximally symmetric 
three-dimensional space given in the Cartesian and the spherical coordinates 
by
\begin{equation}
\gamma_{ij}dx^idx^j=\left(1+{k\over 4}\delta_{mn}x^mx^n\right)^{-2}\delta_{ij}
dx^idx^j={{dr^2}\over{1-kr^2}}+r^2(d\theta^2+\sin^2\theta d\phi^2),
\label{mxsymmet}
\end{equation}
with $k=-1,0,1$ respectively for the three-dimensional space with the 
negative, zero and positive spatial curvature.  

Since $\psi$ is minimally coupled, the Einstein's equations take the 
conventional form:
\begin{equation}
{\cal G}_{MN}={{8\pi G_5}\over\psi}{\cal T}_{MN},
\label{eineq}
\end{equation} 
with the energy-momentum tensor given by
\begin{eqnarray}
{\cal T}_{MN}&=&\partial_M\Phi\partial_N\Phi-{1\over 2}\hat{g}_{MN}\partial_P
\Phi\partial^P\Phi+\hat{g}_{MN}\left({1\over 2}m^2\Phi^2-\Lambda\right)
\cr
& &+\delta^{\mu}_M\delta^{\nu}_N\left[{{\delta(y)}\over b}\left\{
{\cal T}^0_{\mu\nu}+\left(V_0(\Phi)-\sigma_0\right)g_{\mu\nu}\right\}\right.
\cr
& &\left.+{{\delta(y-1/2)}\over b}\left\{{\cal T}^{1/2}_{\mu\nu}
+\left(V_{1/2}(\Phi)-\sigma_{1/2}\right)g_{\mu\nu}\right\}\right],
\label{emtens}
\end{eqnarray}
where the energy-momentum tensors ${\cal T}^{0,1/2}_{\mu\nu}=-{2\over\sqrt{-g}}
{{\delta(\sqrt{-g}{\cal L}_{0,1/2})}\over{\delta g^{\mu\nu}}}$ for  
matter fields on the branes have the following usual perfect-fluid forms:
\begin{equation}
{\cal T}^{0\,\mu}_{\ \ \ \nu}={\rm diag}(-\varrho_0c^2,\wp_0,\wp_0,\wp_0),
\ \ \ \ \ 
{\cal T}^{1/2\,\mu}_{\ \ \ \ \ \nu}={\rm diag}(-\varrho_{1/2}c^2,\wp_{1/2},
\wp_{1/2},\wp_{1/2}).
\label{empf}
\end{equation}
As proposed in Ref.\cite{am}, the time-variable $c$ does not introduce 
corrections to the Einstein tensor ${\cal G}_{MN}$ for the 
bulk metric (\ref{metansat}) in the preferred frame:
\begin{eqnarray}
{\cal G}_{00}&=&3\left[{1\over {c^2}}{\dot{a}\over a}\left({\dot{a}\over a}
+{\dot{b}\over b}\right)-{{n^2}\over b^2}\left\{{a^{\prime}\over a}
\left({a^{\prime}\over a}-{b^{\prime}\over b}\right)+{a^{\prime\prime}\over 
a}\right\}+k{{n^2}\over a^2}\right],
\cr
{\cal G}_{ij}&=&{a^2\over b^2}\left[{a^{\prime}\over a}\left(2{n^{\prime}\over 
n}+{a^{\prime}\over a}\right)-{b^{\prime}\over b}\left({n^{\prime}\over n}+
2{a^{\prime}\over a}\right)+2{a^{\prime\prime}\over a}+{n^{\prime\prime}\over 
n}\right]\gamma_{ij}
\cr
& &+{a^2\over {c^2n^2}}\left[{\dot{a}\over a}\left(2{\dot{n}\over n}
-{\dot{a}\over a}\right)+{\dot{b}\over b}\left({\dot{n}\over n}
-2{\dot{a}\over a}\right)-2{\ddot{a}\over a}-{\ddot{b}\over b}\right]
\gamma_{ij}-k\gamma_{ij},
\cr
{\cal G}_{04}&=&{3\over c}\left[{n^{\prime}\over n}{\dot{a}\over a}+
{a^{\prime}\over a}{\dot{b}\over b}-{\dot{a}^{\prime}\over a}\right],
\cr
{\cal G}_{44}&=&3\left[{a^{\prime}\over a}\left({a^{\prime}\over a}
+{n^{\prime}\over n}\right)-{{b^2}\over{c^2n^2}}\left\{{\dot{a}\over a}
\left({\dot{a}\over a}-{\dot{n}\over n}\right)+{\ddot{a}\over a}\right\}
-k{b^2\over a^2}\right],
\label{eintens}
\end{eqnarray}
where the overdot and the prime respectively denote derivatives w.r.t. 
$t$ and $y$.  Due to the assumption of minimal coupling, the field 
equation for $\Phi$ also takes the usual form:
\begin{eqnarray}
c^{-1}\partial_t\left(n^{-1}a^3bc^{-1}\partial_t\Phi\right)-\partial_y
\left(na^3b^{-1}\partial_y\Phi\right)
\ \ \ \ \ \ \ \ \ \ \ \ \ \ \ \ \ \ \ \ \ \ \ \ \ \ 
\cr
-na^3b\left[m^2\Phi+V^{\prime}_0
{{\delta(y)}\over{b}}+V^{\prime}_{1/2}{{\delta(y-1/2)}\over{b}}\right]=0,
\label{phieq}
\end{eqnarray}
where the primes in the potentials denote derivatives w.r.t. $\Phi$.  

When there is no matter field on the branes, the equations of motion admit 
the static brane solution \cite{rs1} with the metric components given by
\begin{equation}
n(y)=a(y)=e^{-m_0b_0|y|},\ \ \ \ \ \ \ \ 
b(y)=b_0=const.
\label{statsol}
\end{equation}
The brane tensions take the fine-tuned values given by
\begin{equation}
\sigma_0={{3c^4m_0b_0}\over{4\pi G_5}}=-\sigma_{1/2},\ \ \ \ \ \ \ \ \ \ \ 
\Lambda=-{{3c^4m^2_0b^2_0}\over{4\pi G_5}},
\label{fintun}
\end{equation}
from which we see that $m_0$ varies with $t$, if $c$ varies with $t$.  
So, the solution at $y\neq 0$ is not static when $\dot{c}\neq 0$, if the 
the brane tensions $\sigma_{0,1/2}$ are constant, since the parameter 
$m_0$ has to vary with time to keep $\sigma_{0,1/2}$ constant.  In order 
for the solution (\ref{statsol}) to remain static (i.e., for $m_0$ to remain 
constant) while $c$ varies with $t$, the fine-tuned brane tensions 
$\sigma_{0,1/2}$ have to vary with $t$, as specified by Eq. (\ref{fintun}).  

When matter fields are included on the branes, the brane universe undergoes 
cosmological expansion.  In Refs. \cite{cgrt,cf2}, by expanding the metric 
around the static solution (\ref{statsol}) to the linear order in 
$\varrho_{0,1/2}$ and then taking average over $y$, the following effective 
Friedmann equations for the expanding brane universe is obtained:
\begin{equation}
\left({{\dot{a}}\over{a}}\right)^2={{8\pi G_4}\over 3}\left(\varrho_0+
\varrho_{1/2}\Omega^4_0\right)-{{kc^2}\over{a^2}},
\label{frdeq1}
\end{equation}
\begin{equation}
{\ddot{a}\over a}=-{{4\pi G_4}\over 3}\left[\varrho_0+\varrho_{1/2}\Omega^4_0+
{3\over c^2}\left(\wp_0+\wp_{1/2}\Omega^4_0\right)\right],
\label{frdeq2}
\end{equation}
where
\begin{equation}
\Omega_0\equiv e^{-m_0b_0/2},\ \ \ \ \ \ \ \ 
G_4={{m_0G_5}\over{1-\Omega^2_0}}.
\label{defs}
\end{equation}
Note, $G_4$ also varies with $t$, because of the dependence of $m_0$ and 
$\Omega_0$ on $c$.  From the above effective Friedmann equations, we 
obtain the following generalized conservation equation:
\begin{equation}
\dot{\varrho}+3({{\wp}\over{c^2}}+\varrho){\dot{a}\over a}=-\varrho
{\dot{G}_4\over G_4}+{{3kc\dot{c}}\over{4\pi G_4 a^2}},
\label{ncsveq}
\end{equation}
where
\begin{equation}
\varrho\equiv\varrho_0+\varrho_{1/2}\Omega^4_0,\ \ \ \ \ \ \ \ \ \ \ 
\wp\equiv\wp_0+\wp_{1/2}\Omega^4_0.
\label{denpredef}
\end{equation}
So, while $c$ and/or $G_4$ vary with time, mass is not conserved, 
namely that matter is created or taken out of the brane universe.  

In section 2.1 and 2.2, we study the conditions under which the flatness 
problem and the cosmological constant problem can be resolved by assuming 
suddenly changing $c$ at some critical time.  In section 2.3, we consider 
the case in which $c$ gradually changes as a power law in the scale factor 
$a$.

\subsection{The flatness problem}

In this subsection, we illustrate how the flatness problem can be 
resolved by assuming the time-varying $c$.  As the Friedmann equations 
(\ref{frdeq1},\ref{frdeq2}) for the brane universe have the same form as 
those of the standard cosmology, the discussion is along the same line as 
Ref. \cite{bar1}.  The critical density $\varrho_c$, the mass density 
corresponding to $k=0$ for a given $\dot{a}/a$, of the brane universe is 
given by
\begin{equation}
\varrho_c={3\over{8\pi G_4}}\left({\dot{a}\over a}\right)^2.
\label{critden}
\end{equation}
We define the deviation of $\varrho$ from $\varrho_c$ as 
$\epsilon\equiv\varrho/\varrho_c-1$.  So, $\epsilon<0$, $\epsilon=0$, and 
$\epsilon>0$ cases respectively correspond to the open ($k=-1$), flat 
($k=0$) and closed ($k=1$) universes.  Differentiating $\epsilon$ w.r.t. $t$, 
we have 
\begin{equation}
\dot{\epsilon}=(1+\epsilon)\left({\dot{\varrho}\over\varrho}-
{\dot{\varrho}_c\over\varrho_c}\right).
\label{epsdt}
\end{equation}
We assume that the brane matter satisfies the equation of state of 
the form:
\begin{equation}
\wp=w\varrho c^2,
\label{eqst}
\end{equation}
with a constant $w$.  Then, making use of Eqs. 
(\ref{frdeq1},\ref{frdeq2},\ref{ncsveq}), we have
\begin{eqnarray}
{\dot{\varrho}\over\varrho}&=&-3{\dot{a}\over a}(1+w)-{\dot{G}_4\over G_4}
+2{\dot{c}\over c}{\epsilon\over{1+\epsilon}},
\cr
{\dot{\varrho}_c\over\varrho_c}&=&-{\dot{a}\over a}[2+(1+\epsilon)
(1+3w)]-{\dot{G}_4\over G_4}.
\label{dotrhos}
\end{eqnarray}
Substituting these into Eq. (\ref{epsdt}), we have
\begin{equation}
\dot{\epsilon}=(1+\epsilon)\epsilon{\dot{a}\over a}(1+3w)
+2{\dot{c}\over c}\epsilon.
\label{epdotf}
\end{equation}
So, during the SBB evolution (with $\dot{c}=0$), any ordinary matter field 
satisfying the strong energy condition $1+3w>0$ drives $\epsilon$ away from 
zero.  The fact that the value of $\epsilon$ observed in our present universe 
is close to zero (Cf. see Refs. \cite{ber,han,bal} for recent 
observational data) indicates that $\varrho$ at early universe has to be 
remarkably close to $\varrho_c$ to avoid too much deviation from $\varrho_c$ 
at present time, the so-called flatness problem of the SBB model.  The 
inflationary cosmology solves the flatness problems, as can be seen from 
Eq. (\ref{epdotf}):  Since the inflaton scalar violates the strong energy 
condition, $\epsilon$ is driven towards zero during the inflationary era.  
The VSL models can also solve the flatness problem, if $c$ decreases rapid 
enough:  If $|\dot{c}/c|\gg\dot{a}/a$, then the first term on the RHS of Eq. 
(\ref{epdotf}) can be neglected.  So, while $c$ decreases, $\epsilon$ 
is rapidly driven to zero.  Just like the inflationary models, only 
the flat universe ($k=0$) is stable while $c$ decreases.

\subsection{The cosmological constant problem}

We begin by illustrating the cosmological constant problem in the standard 
cosmology and its resolution by the VSL model.  When the contribution 
$\varrho_{\Lambda}=\Lambda c^2/(8\pi G_4)$ of the cosmological constant 
$\Lambda$ is included in $\varrho$, i.e. $\varrho=\varrho_m+\varrho_{\Lambda}$ 
where $\varrho_m$ is the mass density of normal matter, then the generalized 
conservation equation (\ref{ncsveq}) is modified to
\begin{equation}
\dot{\varrho}_m+3{\dot{a}\over a}\left(\varrho_m+{\varrho_m\over c^2}
\right)=-\dot{\varrho}_\Lambda-\varrho{\dot{G}_4\over G_4}+
{{3kc\dot{c}}\over{4\pi G_4a^2}}.
\label{mgenceq}
\end{equation}
If we define $\epsilon_\Lambda\equiv\varrho_\Lambda/\varrho_m$, then 
its derivative w.r.t. $t$ is
\begin{equation}
\dot{\epsilon}_\Lambda=\epsilon_\Lambda\left({\dot{\varrho}_\Lambda
\over\varrho_\Lambda}-{\dot{\varrho}_m\over\varrho_m}\right).
\label{epldt}
\end{equation}
From the definition of $\varrho_\Lambda$, we have
\begin{equation}
{\dot{\varrho}_\Lambda\over\varrho_\Lambda}=2{\dot{c}\over c}-
{\dot{G}_4\over G_4},
\label{drrhl}
\end{equation}
and from Eq. (\ref{mgenceq}), along with the Friedmann equations, we 
obtain
\begin{equation}
{\dot{\varrho}_m\over\varrho_m}=-3{\dot{a}\over a}(1+w)-2{\dot{c}\over 
c}{\varrho_c\over\varrho_m}+2{\dot{c}\over c}-{\dot{G}_4
\over G_4}.
\label{drrhm}
\end{equation}
So, Eq. (\ref{epldt}) takes the form:
\begin{equation}
\dot{\epsilon}_\Lambda=\epsilon_\Lambda\left[3{\dot{a}\over a}(1+w)+
2{\dot{c}\over c}{{1+\epsilon_\Lambda}\over{1+\epsilon}}\right].
\label{epldtf}
\end{equation}
Thus, in the SBB model (with $\dot{c}=0$), any matter satisfying $w>-1$ 
drives $\epsilon_\Lambda$ away from zero as the universe expands, leading 
to the value of the cosmological constant many orders of magnitude larger 
than the presently observed value in our universe.  So, during the initial 
period of the cosmological evolution, $\epsilon_\Lambda$ had to be 
fine-tuned to take extremely small value, the so called cosmological 
constant problem in cosmology.  (To make matters worse, theoretical estimates 
of various contributions rather lead to the cosmological constant of the 
order of the Planck scale.)  So, it might help, if there exist some mechanisms 
that drive $\epsilon_\Lambda$ towards zero.  The inflationary models cannot 
solve the problem, because the violation of the strong energy condition is 
not enough to make $w+1$ negative.  However, the VSL model can solve the 
cosmological constant problem, if $|\dot{c}/c|\gg\dot{a}/a$.  Then, we can 
neglect the first term on the RHS of Eq. (\ref{epldtf}), so 
$\epsilon_\Lambda$ is rapidly driven to zero while $c$ decreases to the 
present value of the speed of light.  In this process, vacuum energy 
density is discharged into ordinary matter.  

We now consider the case of the brane world cosmology.  Unlike the case of 
the standard cosmology, the mass density satisfying $\varrho=-\wp/c^2$ is not 
directly related to the cosmological constant of the brane universe, but 
rather to the brane tension.  However, if we assume the brane tensions to 
have initially taken the fine-tuned values
\footnote{Indeed, in the equations under consideration, i.e. Eqs. 
(\ref{frdeq1},\ref{frdeq2},\ref{ncsveq}), the fine-tuned values for the 
brane tensions $\sigma_{0,1/2}$ are taken into account, and $\varrho$ and 
$\wp$ in these equations do not include contributions from $\sigma_{0,1/2}$.} 
(\ref{fintun}) due to, for example, the supersymmetry embedding of the brane 
world scenario, then by identifying the nonzero mass density 
$\varrho_{\delta\sigma}$ satisfying $\varrho_{\delta\sigma}=
-\wp_{\delta\sigma}/c^2$ as being due to the (quantum) correction 
$\delta\sigma$ to the brane tensions (after the SUSY breaking) we can 
regard nonzero $\varrho_{\delta\sigma}$ as the source of 
nonzero effective cosmological constant in the brane universe.  The total 
mass density of the brane universe is then given by $\varrho=\varrho_m+
\varrho_{\delta\sigma}$, where $\varrho_m$ is the mass density of the 
ordinary matter on the brane and $\varrho_{\delta\sigma}=\delta\sigma/c^2$.  
So, the generalized conservation equation (\ref{ncsveq}) is modified to
\begin{equation}
\dot{\varrho}_m+3{\dot{a}\over a}\left(\varrho_m+{\wp_m\over c^2}
\right)=-\dot{\varrho}_{\delta\sigma}-\varrho{\dot{G}_4\over G_4}+
{{3kc\dot{c}}\over{4\pi G_4a^2}}.
\label{mgenceq2}
\end{equation}
Defining $\epsilon_{\delta\sigma}\equiv\varrho_{\delta\sigma}/\varrho_m$, we 
obtain
\begin{equation}
\dot{\epsilon}_{\delta\sigma}=\epsilon_{\delta\sigma}
\left({\dot{\varrho}_{\delta\sigma}\over\varrho_{\delta\sigma}}
-{\dot{\varrho}_m\over\varrho_m}\right).
\label{epldt2}
\end{equation}
By using the Friedmann equations, the generalized conservation equation and 
the definition of $\epsilon_{\delta\sigma}$, we obtain
\begin{equation}
{\dot{\varrho}_{\delta\sigma}\over\varrho_{\delta\sigma}}=
-2{\dot{c}\over c},
\label{dtrh1}
\end{equation}
\begin{equation}
{\dot{\varrho}_m\over\varrho_m}=-3{\dot{a}\over a}(1+w)-2{\dot{c}\over c}
{\varrho_c\over\varrho_m}+2{\dot{c}\over c}{{\varrho+\varrho_{\delta\sigma}}
\over\varrho_m}-{\varrho\over\varrho_m}{\dot{G}_4\over G_4}.
\label{dtrh2}
\end{equation}
So, Eq. (\ref{epldt2}) takes the form:
\begin{equation}
\dot{\epsilon}_{\delta\sigma}=\epsilon_{\delta\sigma}\left[3{\dot{a}\over a}
(1+w)+2{\dot{c}\over c}{{1+\epsilon_{\delta\sigma}}\over{1+\epsilon}}+
\left({\dot{G}_4\over G_4}-4{\dot{c}\over c}\right)(1+
\epsilon_{\delta\sigma})\right].
\label{epdtf2}
\end{equation}
Note, $G_4$ is given by Eq. (\ref{defs}).  We also assume that $\dot{G}_5
\neq 0$.  Then, we obtain
\begin{equation}
{\dot{G}_4\over G_4}=-4{\dot{c}\over c}+2{\dot{G}_5\over G_5}
+{{16\pi G_5\sigma_0\Omega^2_0}\over{3c^4(1-\Omega^2_0)}}{\dot{c}\over c}
-{{4\pi G_5\sigma_0\Omega^2_0}\over{3c^4(1-\Omega^2_0)}}{\dot{G}_5\over G_5}.
\label{gdt}
\end{equation}
As before, the first term on the RHS of Eq. (\ref{epdtf2}) can be neglected, 
if we assume $|\dot{c}/c|, |\dot{G}_5/G_5|\gg \dot{a}/a$.  The coefficients 
of $\dot{c}/c$ and $\dot{G}_5/G_5$ for the last two terms on the RHS of 
Eq. (\ref{gdt}) are of the order $10^{-15}$ with values of quantities assumed 
in the RS model \cite{rs1,rs2,rs3}, so the last two terms can be neglected 
compared to the first two terms.  Then, Eq. (\ref{epdtf2}) becomes
\begin{equation}
\dot{\epsilon}_{\delta\sigma}\approx\epsilon_{\delta\sigma}\left(2{\dot{G}_5
\over G_5}-{{6+8\epsilon}\over{1+\epsilon}}{\dot{c}\over c}\right)
(1+\epsilon_{\delta\sigma}).
\label{epdtf22}
\end{equation}
First, we see from Eq. (\ref{epdtf2}) that corrections to the fine-tuned 
brane tensions (\ref{fintun}) will grow fast as the brane universe expands, 
if we assume $\dot{c}=0=\dot{G}_5$.  Therefore, in the brane world 
cosmologies the corrections to the brane tensions at an early time have to 
be extremely small or there has to exist some mechanism that suppresses 
growth of corrections to the brane tensions, in order to avoid contradiction 
with small cosmological constant observed in our present universe.  Second, 
from Eq. (\ref{epdtf22}) we see that, unlike the case of the VSL standard 
cosmology, assuming $\dot{c}<0$ with the constant (five-dimensional) Newton's 
constant is not enough for resolving the cosmological constant problem.  In 
fact, the fast decreasing $c$ will make the corrections to the brane tensions 
grow extremely fast, leading to very large value of cosmological constant in 
brane universe.  In order to resolve the cosmological constant problem, we 
have to additionally assume that $G_5$ decreases faster than $c$:
\begin{equation}
2{\dot{G}_5\over G_5}<{{6+8\epsilon}\over{1+\epsilon}}{\dot{c}\over c}.
\label{gcdec}
\end{equation}
When this condition is satisfied, the correction $\delta\sigma=\delta\sigma_0
+\Omega^4_0\delta\sigma_{1/2}$ to the fine-tuned brane tensions 
$\sigma_{0,1/2}$ is converted into the conventional matter while $c$ and 
$G_5$ decrease, thereby the brane tensions are pushed back to the fine-tuned 
values (\ref{fintun}).

\subsection{Gradually changing speed of light}

In this subsection, we examine the case in which $c$ varies as a power law 
in the scale factor $a$, as was first considered in Ref. \cite{bar1}.  Since 
the issue on the flatness problem is along the same line as the standard 
cosmology case, we concentrate on the cosmological constant problem.  
We can rewrite Eq. (\ref{mgenceq2}) with $\varrho_m=w\wp_mc^2$ in the 
following form:
\begin{equation}
\left(G_4\varrho_ma^{3(w+1)}\right)^.=G_4\varrho_{\delta\sigma}a^{3(w+1)}
\left(2{\dot{c}\over c}-{\dot{G}_4\over G_4}\right)+{{3kc\dot{c}a^{3w+1}}
\over{4\pi}}.
\label{mgenceq22}
\end{equation}
By applying the same approximation used in the previous subsection, namely 
that the last two terms on the RHS of Eq. (\ref{gdt}) can be 
neglected and similarly $\Omega_0\approx 0$ in Eq. (\ref{defs}), i.e. 
$G_4\approx m_0G_5$, we can approximate Eq. (\ref{mgenceq22}) to the 
following form:
\begin{equation}
\left(G_4\varrho_ma^{3(w+1)}\right)^.\approx{{8\pi\sigma_0\delta\sigma G^2_5
a^{3(w+1)}}\over{3b_0c^6}}\left(3{\dot{c}\over c}-{\dot{G}_5\over G_5}\right)
+{{3kc\dot{c}a^{3w+1}}\over{4\pi}}.
\label{mgenceq23}
\end{equation}
Assuming that $c$ and $G_5$ vary gradually like $c=c_0a^{\bf n}$ and 
$G_5=\tilde{G}_5a^{\sf q}$ with constant $c_0$, $\tilde{G}_5$, ${\sf n}$ and 
${\sf q}$, we can integrate Eq. (\ref{mgenceq23}) to obtain
\begin{eqnarray}
G_4\varrho_ma^{3(w+1)}&=&{{3{\sf n}-{\sf q}}\over{3w+2{\sf q}-6{\sf n}+3}}
{{8\pi\tilde{G}^2_5\sigma_0\delta\sigma}\over{3b_0c^6_0}}a^{3w+2{\sf q}-
6{\sf n}+3}
\cr
& &+{{3{\sf n}kc^2_0}\over{4\pi(3w+2{\sf n}+1)}}a^{3w+2{\sf n}+1}+B,
\label{mgenceqsol}
\end{eqnarray}
where $B$ is an integration constant.  Substituting this into Eq. 
(\ref{frdeq1}) with $\varrho=\varrho_m+\varrho_{\delta\sigma}$, we have
\begin{equation}
\left({\dot{a}\over a}\right)^2={{8\pi}\over 3}{B\over{a^{3w+3}}}+
{{32(w+1)\pi^2}\over{3(3w+2{\sf q}-6{\sf n}+3)}}{{\tilde{G}^2_5\sigma_0\delta
\sigma}\over{b_0c^6_0}}a^{2{\sf q}-6{\sf n}}-{{3w+2{\sf n}-1}\over{3w+
2{\sf n}+1}}kc^2_0a^{2{\sf n}-2},
\label{fredgv}
\end{equation}
where the second and the third terms on the RHS respectively correspond to 
the cosmological constant and the curvature terms.  
The ratio of the matter energy density and the corrections to the brane 
tensions is given by
\begin{equation}
{\varrho_m\over\varrho_{\delta\sigma}}={{2b_0c^6_0B}\over{\tilde{G}^2_5
\sigma_0\delta\sigma}}a^{-3w+6{\sf n}-2{\sf q}-3}+{{9{\sf n}kc^8_0b_0}\over
{16\pi^2\tilde{G}^2_5\sigma_0\delta\sigma(3w+2{\sf n}+1)}}a^{8{\sf n}-2{\sf q}
+1}+{{6{\sf n}-2{\sf q}}\over{3w-6{\sf n}+2{\sf q}+3}}.
\label{ratio}
\end{equation}

Note, in the case of the VSL brane world cosmology, the cosmological 
constant term in the Friedmann equation behaves as $\sim a^{2q-6n}$ 
with the {\it negative} coefficient for the ${\sf n}$ term in the index, 
whereas that of the VSL standard cosmology behaves as $\sim a^{2n}$ 
\cite{bar1,bar3}.  So, we have to further assume that the Newton's constant 
$G_5$ gradually decreases fast enough in order for the $B$ term dominates the 
cosmological constant term as $a$ increases.  The condition for this to 
happen is given by
\begin{equation}
3w-6{\sf n}+2{\sf q}-3<0.
\label{lambcond}
\end{equation}
As in the VSL standard cosmology case \cite{bar1,bar3}, the condition for 
the curvature term to be dominated by the $B$ term as $a$ increases is given by
\begin{equation}
3w+2{\sf n}+1<0.
\label{curvcond}
\end{equation}
So, unlike the VSL standard cosmology case, resolving the flatness problem 
does not necessary guarantees resolving the cosmological constant problem.  

Next, we consider the quasi cosmological constant problem.  The quasi 
cosmological constant problem stems from the fact that the cosmological 
constant observed in our present universe is not exactly zero but has a 
small positive value.  One of the appealing features of the VSL models is 
that it resolves not only the cosmological constant problem but also the 
quasi cosmological constant problem.  As can be seen in Eq. (\ref{ratio}), 
for a suitable choice of ${\sf n}$ and ${\sf q}$ the ratio of the matter 
density and the corrections to the brane tensions (therefore the vacuum 
energy density) approaches a constant value $(6{\sf n}-2{\sf q})/(3w-6{\sf 
n}+2{\sf q}+3)$ as $a$ increases.  When Eq. (\ref{lambcond}) is satisfied, 
just like the VSL standard cosmology case, there is no solution to the 
quasi cosmological constant problem, since the ratio (\ref{ratio}) 
approaches infinity (i.e. the cosmological constant approaches zero) as 
$a$ increases.  However, unlike the VSL standard cosmology case, we cannot 
resolve the quasi cosmological constant problem just by assuming $3w-6{\sf n}
+2{\sf q}-3>0$:  We have to further assume that $8{\sf n}-2{\sf q}+1<0$ so 
that the second term on the RHS of Eq. (\ref{ratio}) approaches zero as $a$ 
increases.   Also, in order for the ratio to approach a {\it positive} value, 
we have to additionally assume that $3{\sf n}>{\sf q}$.  Note, the condition 
$3{\sf n}>{\sf q}$ also corresponds to the condition that the cosmological 
constant term on the RHS of Eq. (\ref{fredgv}) approaches zero as $a$ 
increases.  When these conditions are satisfied, the corrections 
$\delta\sigma_{0,1/2}$ to the fine-tuned brane tensions will reach some 
equilibrium values which give rise to a small positive cosmological constant 
in the brane universe, as $a$ increases.

\section{The VSL Cosmology in the RS2 Model}

In this section, we consider the VSL cosmology in the RS model with 
noncompact extra dimension.  In the preferred frame in which the principle 
of minimal coupling holds, the action takes the form:
\begin{equation}
S=\int d^5x\left[\sqrt{-\hat{g}}\left({\psi\over{16\pi G_5}}{\cal R}-\Lambda
\right)+{\cal L}_{\psi}\right]
+\int d^4x\sqrt{-g}\left[{\cal L}_{mat}-\sigma\right],
\label{sgbrnact2}
\end{equation}
where ${\cal L}_{mat}$ is the Lagrangian density for matter fields confined 
on the brane with tension $\sigma$ at $y=0$.  So, the energy-momentum 
tensor is given by
\begin{equation}
{\cal T}_{MN}=-\hat{g}_{MN}\Lambda+\delta^{\mu}_M\delta^{\nu}_N\left(
{\cal T}^{mat}_{\mu\nu}-g_{\mu\nu}\sigma\right){{\delta(y)}\over b},
\label{emtens2}
\end{equation}
with the energy-momentum tensor ${\cal T}^{mat}_{\mu\nu}=-{2\over
\sqrt{-g}}{{\delta(\sqrt{-g}{\cal L}_{mat})}\over{\delta g^{\mu\nu}}}$ 
for the brane matter fields having the usual perfect fluid form:
\begin{equation}
{\cal T}^{mat\,\mu}_{\ \ \ \ \ \ \nu}={\rm diag}(-\varrho c^2,\wp,\wp,\wp).
\label{empf2}
\end{equation}
The Einstein's equations (\ref{eineq}) therefore take the form:
\begin{equation}
{3\over {c^2n^2}}{\dot{a}\over a}\left({\dot{a}\over a}+{\dot{b}\over b}
\right)-{3\over b^2}\left[{a^{\prime}\over a}\left({a^{\prime}\over a}-
{b^{\prime}\over b}\right)+{a^{\prime\prime}\over a}\right]+{{3k}\over a^2}= 
{{8\pi G_5}\over c^4}\left[\Lambda+(\sigma+\varrho c^2){{\delta(y)}\over b}
\right],
\label{expeineqs1}
\end{equation}
\begin{eqnarray}
{1\over b^2}\left[{a^{\prime}\over a}\left(2{n^{\prime}\over n}+{a^{\prime}
\over a}\right)-{b^{\prime}\over b}\left({n^{\prime}\over n}+2{a^{\prime}
\over a}\right)+2{a^{\prime\prime}\over a}+{n^{\prime\prime}\over n}\right]+
\ \ \ \ \ \ \ \ \ \ \ \ \ \ \ \ \ \ \ \ \ \ \ \ \ \ \ \ \ \ \ \ \ \ \ \ \ \ 
\ \ \ 
\cr
{1\over {c^2n^2}}\left[{\dot{a}\over a}\left(2{\dot{n}\over n}
-{\dot{a}\over a}\right)+{\dot{b}\over b}\left({\dot{n}\over n}
-2{\dot{a}\over a}\right)-2{\ddot{a}\over a}-{\ddot{b}\over b}\right]
-{k\over a^2}=
-{{8\pi G_5}\over c^4}\left[\Lambda+(\sigma-\wp){{\delta(y)}\over b}\right],
\label{expeineqs2}
\end{eqnarray}
\begin{equation}
{n^{\prime}\over n}{\dot{a}\over a}+{a^{\prime}\over a}{\dot{b}\over b}
-{\dot{a}^{\prime}\over a}=0,
\label{expeineqs3}
\end{equation}
\begin{equation}
{3\over b^2}{a^{\prime}\over a}\left({a^{\prime}\over a}+{n^{\prime}\over n}
\right)-{3\over{c^2n^2}}\left[{\dot{a}\over a}\left({\dot{a}\over a}-{\dot{n}
\over n}\right)+{\ddot{a}\over a}\right]-{{3k}\over 
a^2}=-{{8\pi G_5}\over c^4}\Lambda.
\label{expeineqs4}
\end{equation}  

The effective four-dimensional equations of motion on the three-brane 
worldvolume can be obtained \cite{bdl} by taking the jumps and the mean values 
of the above Einstein's equations across $y=0$ and then applying the boundary 
conditions on the first derivatives of the metric components due to 
the $\delta$-function singularity on the brane.   We consider the 
solution invariant under the ${\bf Z}_2$-symmetry $y\to -y$.  
We define $t$ to be the cosmic time for the brane universe, i.e. $n(t,0)=1$.  
We also define the $y$-coordinate to be proportional to the proper distance 
along the $y$-direction with $b$ being the constant of proportionality, 
namely $b^{\prime}=0$.  We further assume that the radius of the extra space 
is stabilized, i.e. $\dot{b}=0$.  Making use of these assumptions, we define 
the $y$-coordinate such that $b=1$.  The resulting effective Friedmann 
equations have the forms:
\begin{equation}
\left({\dot{a}_0\over a_0}\right)^2={{32\pi^2 G^2_5}\over{9c^6}}
(\varrho^2c^4+2\sigma\varrho c^2)+{{c^2a^{\prime\prime}_{R\,0}}\over{a_0}}+
{{8\pi G_5}\over{3c^2}}\left(\Lambda+{{4\pi G_5}\over{3c^4}}\sigma^2\right)
-{{kc^2}\over a^2_0},
\label{00eq}
\end{equation}
\begin{equation}
{\ddot{a}_0\over a_0}=-{{32\pi^2 G^2_5}\over{9c^6}}(2\varrho^2c^4+
\sigma\varrho c^2+3\sigma\wp+3\varrho\wp c^2)-{{c^2a^{\prime\prime}_{R\,0}}
\over{a_0}}+{{32\pi^2 G^2_5}\over{9c^6}}\sigma^2,
\label{aceq}
\end{equation}
where $a^{\prime\prime}_R$ denotes the regular part of $a^{\prime\prime}$ 
and the subscript $0$ denotes the quantities evaluated at $y=0$, e.g. 
$a_0(t)\equiv a(t,0)$.  

The $a^{\prime\prime}_{R\,0}$-terms (called ``dark radiation'' terms) in the 
above Friedmann equations originate from the Weyl tensor of the bulk and 
thus describe the backreaction of the bulk gravitational degrees of freedom 
on the brane \cite{bdl,bdel,ftw,muk,ida}.  These terms can be evaluated by 
solving $a$ as a function of $y$ from the following equation obtained from 
the diagonal component Einstein's equations:
\begin{equation}
3{{a^{\prime\prime}_R}\over{a}}+{{n^{\prime\prime}_R}\over{n}}=
-{{16\pi G_5}\over{3c^4}}\Lambda,
\label{anbulk}
\end{equation}
supplemented by the following relation obtained from the $(0,4)$-component 
Einstein's equation with the assumed $\dot{b}=0$ condition:
\begin{equation}
n(t,y)=\lambda(t)\dot{a}(t,y),
\label{anrel}
\end{equation}
where $\lambda(t)$ is an arbitrary function of $t$.  The resulting expression 
is
\begin{equation}
a^{\prime\prime}_R={C\over a^3}-{{4\pi G_5}\over{3c^4}}\Lambda a
-{{4\pi G_5}\over{3a^3}}\Lambda\int c^{-5}\dot{c}a^4\,dt,
\label{appsol}
\end{equation}
where $C$ is an integration constant.  For the case where $c$ suddenly 
jumps from $c_-$ to $c_+$ at a critical time $t_c$, the integrand in 
Eq. (\ref{appsol}) becomes  a $\delta$-function $c^{-5}\dot{c}=
{{c^4_+-c^4_-}\over{4c^4_+c^4_-}}\delta(t-t_c)$, so the integral term 
can be easily evaluated 
\footnote{For the $c(t)=c_0a^{\sf n}$ case, the integral can also be easily 
integrated to be $\int c^{-5}\dot{c}a^4\,dt=-{{{\sf n}c^{-4}_0}\over
{4({\sf n}-1)}}a^{-4({\sf n}-1)}$.}.  

We assume that $\sigma$ is fine-tuned as in Eq. (\ref{fintun}), and 
$\sigma\gg \varrho c^2,\wp$ \cite{cgk,cgs}.  To the leading order, the 
Friedmann equations (\ref{00eq},\ref{aceq}) then take the following forms:
\begin{equation}
\left({\dot{a}_0\over a_0}\right)^2={{8\pi G_4}\over 3}\varrho
+{{Cc^2}\over{a^4_0}}-{{kc^2}\over a^2_0},
\label{00eqdel}
\end{equation}
\begin{equation}
{\ddot{a}_0\over a_0}=-{{4\pi G_4}\over 3}(\varrho+3{\wp\over c^2})
-{{Cc^2}\over{a^4_0}},
\label{aceqdel}
\end{equation}
where the effective four-dimensional Newton's constant is given by
\begin{equation}
G_4={{8\pi G^2_5\sigma}\over{3c^4}},
\label{4dg}
\end{equation}
and we have absorbed the coefficient of the last term in Eq. (\ref{appsol}) 
evaluated at $y=0$ into the integration constant $C$.  From these effective 
Friedmann equations, we obtain the following generalized energy 
conservation equation:
\begin{equation}
\dot{\varrho}+3{\dot{a}_0\over a_0}\left(\varrho+{\wp\over c^2}\right)=
-{\dot{G}_4\over G_4}\varrho+{{3kc\dot{c}}\over{4\pi G_4a^2_0}}-
{{3Cc\dot{c}}\over{4\pi G_4a^4_0}}.
\label{ecnseq}
\end{equation}
So, in the case of VSL cosmology in the RS2 model, the dark radiation terms 
make additional contribution to nonconservation of mass density in the brane 
universe.  The constant $C$ can take an arbitrary value, depending on the 
geometry in the bulk, but its upper limit can be constrained by 
nucleosynthesis \cite{bdel}.  In Ref. \cite{msm}, it is shown by considering 
the brane world cosmology on a domain wall moving in the bulk of the AdS 
Schwarzschild space that $C$ can be identified as the mass of the AdS black 
hole.  The requirement of existence of regular horizon for the bulk spacetime 
gives rise to the lower bound for $C$.

\subsection{The flatness problem}

In this subsection, we examine the condition under which the flatness problem 
can be resolved in the VSL cosmology of the RS2 model.  From Eq. 
(\ref{00eqdel}) we see that the critical density is now given by
\begin{equation}
\varrho_c={3\over{8\pi G_4}}\left[\left({\dot{a}_0\over a_0}\right)^2-
{{Cc^2}\over a^4_0}\right].
\label{critden2}
\end{equation}
So, for a sufficiently large value of $C$, the critical mass density becomes 
negative, in which case the flatness problem gets worse as we will see.  
The time derivative of $\epsilon=\varrho/\varrho_c-1$ is still given 
by Eq. (\ref{epsdt}).  However, the dark radiation terms modify 
$\dot{\varrho}/\varrho$ and $\dot{\varrho}_c/\varrho_c$ in the following way
\begin{eqnarray}
{\dot{\varrho}\over\varrho}&=&-3{\dot{a}_0\over a_0}(1+w)-{\dot{G}_4\over G_4}
+2{\dot{c}\over c}\left[{\epsilon\over{1+\epsilon}}-{{3Cc^2}\over
{8\pi G_4a^4_0\varrho}}\right],
\cr
{\dot{\varrho}_c\over\varrho_c}&=&-{\dot{a}_0\over a_0}[2+(1+\epsilon)
(1+3w)]-{\dot{G}_4\over G_4}-{{3Cc\dot{c}}\over{4\pi G_4a^4_0\varrho_c}}.
\label{vardots}
\end{eqnarray}
Substituting these into Eq. (\ref{epsdt}), we obtain
\begin{eqnarray}
\dot{\epsilon}&=&(1+\epsilon)\epsilon{\dot{a}_0\over a_0}(1+3w)
+2{\dot{c}\over c}\epsilon+{{3Cc\dot{c}}\over{4\pi G_4a^4_0\varrho_c}}
\epsilon
\cr
&=&(1+\epsilon)\epsilon{\dot{a}_0\over a_0}(1+3w)+
{{2\left({\dot{a}_0\over a_0}\right)^2}\over{\left({\dot{a}_0\over a_0}
\right)^2-{{Cc^2}\over{a^4_0}}}}{\dot{c}\over c}\epsilon,
\label{depeq}
\end{eqnarray}
where Eq. (\ref{critden2}) is used in the second equality.  
So, when $C<0$, the flatness problem can be resolved by assuming 
rapid enough decreasing $c$, just as in the RS1 case.  
However, when $C\geq 0$, the flatness problem can be resolved by assuming 
$|\dot{c}/c|\gg\dot{a}_0/a_0$, provided $C$ satisfies $C<(a_0\dot{a}_0/c)^2$.  
Too large positive value for $C$, leading to negative $\varrho_c$, will 
drive $\epsilon$ away from zero fast.  This upper limit on $C$ is compatible 
with the upper limit imposed by the nucleosythesis constraint \cite{bdel} 
and the fact that $C$ should not be large in order not to give a large 
contribution to the current expansion of the universe.

\subsection{The cosmological constant problem}

We now examine the condition under which the cosmological constant problem 
can be resolved in the VSL cosmology of the RS2 model.  Separating the 
contribution to the mass density into that from the ordinary matter and that 
from the correction to the brane tension, i.e. $\varrho=\varrho_m+
\varrho_{\delta\sigma}$, we can rewrite the generalized conservation 
equation (\ref{ecnseq}) in the following form:
\begin{equation}
\dot{\varrho}_m+3{\dot{a}_0\over a_0}\left(\varrho_m+{\wp_m\over c^2}\right)=
-\dot{\varrho}_{\delta\sigma}-{\dot{G}_4\over G_4}\varrho+{{3kc\dot{c}}
\over{4\pi G_4a^2_0}}-{{3Cc\dot{c}}\over{4\pi G_4a^4_0}}.
\label{ecnseq2}
\end{equation}
$\dot{\varrho_{\delta\sigma}}/\varrho_{\delta\sigma}$ has the same form 
(\ref{dtrh1}) as the RS1 case, but $\dot{\varrho}_m/\varrho_m$ receives an 
additional contribution from the dark radiation terms:
\begin{equation}
{\dot{\varrho}_m\over\varrho_m}=-3{\dot{a}\over a}(1+w)-2{\dot{c}\over c}
{\varrho_c\over\varrho_m}+2{\dot{c}\over c}{{\varrho+\varrho_{\delta\sigma}}
\over\varrho_m}-{\varrho\over\varrho_m}{\dot{G}_4\over G_4}-{{3Cc\dot{c}}
\over{4\pi G_4a^4_0\varrho_m}}.
\label{dtrhr2}
\end{equation}
Eq. (\ref{epldt2}) therefore takes the form:
\begin{equation}
\dot{\epsilon}_{\delta\sigma}=\epsilon_{\delta\sigma}\left[3{\dot{a}\over a}
(1+w)+2{\dot{c}\over c}{{1+\epsilon_{\delta\sigma}}\over{1+\epsilon}}+
\left({\dot{G}_4\over G_4}-4{\dot{c}\over c}\right)(1+\epsilon_{\delta\sigma})
+{{3Cc\dot{c}}\over{4\pi G_4a^4_0\varrho_m}}\right].
\label{epdtfr2}
\end{equation}
Making use of the expression for $G_4$ in Eq. (\ref{4dg}) and assuming 
$\dot{G}_5\neq 0$, we can rewrite this as
\begin{equation}
\dot{\epsilon}_{\delta\sigma}=\epsilon_{\delta\sigma}\left[3{\dot{a}\over a}
(1+w)+\left\{2{\dot{G}_5\over G_5}-{{6+8\epsilon}\over{1+\epsilon}}{\dot{c}
\over c}\right\}(1+\epsilon_{\delta\sigma})+{{9Cc^6}\over{32\pi^2G^2_5\sigma 
a^4_0\varrho_m}}{\dot{c}\over c}\right].
\label{epdtfr2f}
\end{equation}
So, in the brane world cosmology of the RS2 model with $\dot{c}=0$ and  
$\dot{G}_5=0$, the correction to the fine-tuned brane tension will 
grow fast as the brane universe expands, unless the correction is extremely 
small.  Unlike the RS1 case considered in the previous section, there is a 
possibility of solving the cosmological constant without assuming $\dot{G}_5
\neq 0$.  Namely, when $\dot{G}_5=0$, the dark radiation term plays a crucial 
role:  The integration constant $C$ has to take large enough positive value 
to be able to suppress the growth of $\epsilon_{\delta\sigma}$ with time.  
Otherwise, we have to further assume rapid enough decreasing $\dot{G}_5$.  
Note, however, that too large positive value for $C$ will drive 
$\epsilon$ away from zero fast, as seen in the previous subsection.  
So, when $\dot{G}_5=0$, $C$ should not take too large or too small positive 
value in order to solve both flatness and cosmological constant problems.

\end{document}